\documentstyle[fleqn,espcrc2]{article}
\font\boldgreek=cmmib10
\mathchardef\mypsi="0920
\def\bfpsi{{\fam=9 \mypsi}\fam=1}
\def\beq{\begin{equation}}
\def\eeq{\end{equation}}
\title{Nucleus-nucleus collisions : what we have learned from the heavy ion program at CERN} 
\author{A. Capella\address{Laboratoire de Physique Th\'eorique et Hautes Energies \\
 Universit\'e Paris XI - b\^atiment 211, 91405 Orsay Cedex,
France}\thanks{Laboratoire associ\'e au Centre National de la Recherche Scientifique
- URA D0063}}

\begin{document}
\begin{abstract}
In the first part, I give a brief description of the quark-gluon plasma search at CERN and of some
experimental results. In the second part, I review a dynamical model of nucleus-nucleus
interactions and propose a physical interpretation of those results. \end{abstract}
\maketitle 
\vspace{1 truecm} 

\section{QUARK-GLUON PLASMA SEARCH} \par

Heavy ion collisions is a very active field in particle physics. At present the maximal energy is
reached at CERN where lead ions are ac\-ce\-le\-ra\-ted to an energy of 160 GeV/c per nucleon. The
next machine will be the RHIC collider, expected to operate in 1999, in which gold or lead ions will
reach a c. of m. energy of $\sqrt{s} =$ 200 GeV. One decade later an energy of $\sqrt{s} =$ 5.5
TeV will be reached at CERN-LHC. The aim of these programs it to produce and detect a new state of
matter called quark matter (QM) or quark-gluon plasma (QGP). Since ordinary matter is composed of
quark and gluons it is natural to think that by heating and/or compressing it, a new phase of
deconfined quarks and gluons can be reached. This idea is quite old and was first explored in the
framework of the bag model. Ho\-we\-ver, it only gained full credibility when lattice calculations
showed that such a phase transition does occur in statistical QCD. \par \vskip 5 truemm

\subsection{Lattice QCD results} \par
In the case of a pure Yang-Mills theory (i.e. QCD without quarks) the phase transition is of first
order and takes place at a critical temperature $T_C \sim$ 150 MeV. In the Big Bang theory the same
transition occurred, in the opposite direction (i.e. from QGP to ordinary nuclear matter), when the
universe was only a few microseconds old. Schematically, the situation is as follows. For $T < T_c$
one heats the system and the temperature increases whereas the ratio $\varepsilon/T^4$, where
$\varepsilon$ is the energy density, remains constant. At $T = T_c$ one can heat the system but the
temperature does not increase (only the latent heat does). Finally, for $T > T_c$ the temperature
increases again and $\varepsilon /T^4$ remains constant. This new cons\-tant is much larger than
the cor\-res\-pon\-ding one for $T < T_c$. This is due to the larger number of degrees of freedom of
a QGP as compared to a pion gas. Using the value of the new constant and the value of $T_c$ one finds
that the phase transition takes place when the density of the system is an order of magnitude larger
than that of ordinary nuclear matter. It is also found that there exists a restoration of chiral
symmetry which takes place at the same temperature $T_c$. \par

When quarks are introduced, the calculations are more complicated and the order of the phase
transition depends on the number of light quark flavors. \par \vskip 5 truemm

\subsection{QGP production} \par
Although QGP could be produced in the case of neutron stars we are interested here in its
production in heavy ion collisions. The first remark is that, in this case, we are a priori
very far from the conditions of thermodynamical e\-qui\-li\-brium assumed in the lattice calculations
in statistical QCD. However, as far as the energy density is concerned, it is plausible that
values of the order required by theory (i.e. 1 GeV/fm$^3$) can be reached in central $Pb$-$Pb$
collisions at present CERN energies. In the Bjorken scenario, the e\-ner\-gy density is given by

\beq
\varepsilon = {[dN/dy]_{y^* \sim 0} <E_T> \over \pi \ R_{P_b}^2 \ \tau_0} \ \ \ , 
\label{1e}
\eeq

\noindent where $dN/dy$ is the number of secondaries per unit rapidity, $<E_T>$ is the average
transverse energy per particle, and $\tau_0$ is the plasma formation time. This time is very
poorly known and there is no clear consensus among experts in the field. Using $\tau_0 =$ 1 fm
together with the experimental values of the other quantities in (\ref{1e}), one gets $\varepsilon
\sim 2 \div 3$ GeV/fm$^3$. \par \vskip 5 truemm

\subsection{QGP signals} \par
Even if QGP is produced, the phenomenon is a transient one, and one has to find experimental
signals of its formation. Although some signals regarding the bulk properties of the production
have been proposed in the past (i.e. behavior of $<p_T>$ versus $dn/dy$), it seems at present that
the best signals correspond to rear events or to electromagnetic processes. I will only discuss
here two of these signals~: strangeness enhancement, and $J/\psi$ suppression. \par

Due to chiral symmetry restoration it is expected that the ratio of strange over non-strange
particles will increase due to a reduction of the production threshold in dense hadronic matter. \par

The suppression of $J/\psi$ production, proposed by Matsui and Satz, is a consequence of color
screening in a plasma (Debye screening) - similar to the screening of electric charge in an
ordinary plasma. This screening produces a modification of the potential of the type

\beq
V(r) = V_0(r) \exp (- r/r_D(T))
\label{2e}
\eeq 

\noindent where $r_D$ is the Debye radius - which decreases when $T$ increases. When $r_D(T)$ is
smaller than the binding radius, the corresponding bound state cannot exist. Of course, at
sufficiently high $T$, there will be a transient melting of all hadrons. The interest of the
$J/\psi$ (as well as other charmonium and bottonium states) is twofold. First, lattice QCD
calculations show that the melting of $J/\psi$ occurs at temperatures $T \sim 1.2 \ T_C$ only
slightly above the critical temperature and, se\-cond, $c$ and $\bar{c}$ being very rear will
combine with light quarks to produce open charmed mesons and therefore, once melted, the $J/\psi$
will not be formed again. 

\subsection{Experimental situation}
The two phenomena described above have been observed experimentally in a very clear way. Indeed it
has been found that the values of the ratios
\beq
{K \over \pi} \quad , \quad {\Lambda \over \pi} \quad , \quad {\bar{\Lambda} \over \pi} \quad ,
\quad {\Xi \over \Lambda} \quad , \cdots
\label{3ea} 
\eeq
\noindent increase between $pp$ or $pA$ and central $AB$ collisions. The NA35 collaboration
claims that there is no increase between $pp$ and $pA$, but errors are still large. \par

$J/\psi$ suppression has been observed by mea\-su\-ring the ratio $J/\psi$ over Drell-Yan (DY). It is
found that this ratio decreases continuously from $pp$ to $pA$ to $AB$ collisions. Furthermore, for
a given $AB$ system, the ratio decreases with increasing centrality. This effect is very strong in
$Pb$-$Pb$ collisions (see Section 2.6 and Fig. 8). \par \vskip 5 truemm

\subsection{Physical interpretation} \par
There are claims in the literature that strangeness enhancement and particularly the enhancement
of multi-strange antibaryons indicates the formation of QGP. Nevertheless, it has also been possible
to explain these data in the framework of string models - at the cost of in\-tro\-du\-cing some new
ingredients such as string fusion and final state rescattering of secondaries (see section 2.5).
Likewise, alternative explanations of the observed $J/\psi$ suppression have been proposed (see
2.6). \par

In the following I will briefly review an independent string model~: the dual parton model (DPM)
and will discuss to what extent the above signals can be understood in this framework, what are
the required modifications of the model and what these heavy ion experiments teach us regarding
the mechanism of multiparticle production. \par \vskip 5 truemm

\section{STRING MODELS AND QGP SI\-GNALS} \par
The dual parton model (DPM) is a dy\-na\-mi\-cal model aimed at describing the mechanism of
multiparticle production in hadronic and nuclear collisions. A very similar model, the quark gluon
string model (QGSM) has been introduced by the ITEP group. These models are based on the large-$N$
expansion of non-perturbative QCD. A comprehensive review can be found in ref. \cite{1r}. A short
description of the physical basis of the model and some of its applications to hadronic collisions
can be found in \cite{2r}. Here I will concentrate on the generalization of the model to nuclear
collisions, and, in particular, on the eventual modifications of the model required to explain the
CERN heavy ion data.  \par 

\subsection{The model} \par
In $pp$ collisions the main contribution consists in a color exchange producing two strings of type
diquark quark (see Fig. 1). The hadronic spectra of each string is obtained by a convolution of
the momentum distribution and fragmentation func\-tions of the constituents at the string ends. Due
to energy-momentum conservation, the longitudinal momentum fractions of the constituents at the
string ends in each hemisphere add up to unity. There are also more complicated configurations,
corresponding to higher order terms in the large-$N$ expansion, involving 4, 6, $\dots$ etc strings,
with sea quarks and antiquarks at their ends. These configurations correspond to multiple inelastic
scattering in an $S$-matrix approach. In $pp$ collisions, these higher order configurations give
small contributions at moderate energies, but their contribution becomes increasingly important with
increasing energy. In proton-nucleus and nucleus-nucleus collisions, these higher-order
contributions are enhanced by trivial combinatorial factors (proportional to $A^{1/3}$ in $pA$
collisions). More precisely, in $pA$ collisions the dominant graph, involving only one struck nucleon
of $A$ and $A - 1$ spectator nucleons, is the same two string diagram of Fig. 1. A double inelastic
collision, involving two struck nucleons of $A$ and $A - 2$ spectators, corresponds to a four
string graph as shown in Fig. 2, and so on. The weights of these contributions can not be computed
in QCD and are taken from the Glauber-Gribov approach.  

\begin{figure}
\vspace{5 truecm}
\caption{Dominant two string component in nucleon-nucleon scattering.} 
\end{figure}

\begin{figure}
\vspace{6 truecm}
\caption{Four string component in nucleon-nucleus scattering for two inelastic
collisions.} 
\end{figure}

One assumption of DPM, which is not motivated by the large-$N$ expansion, is that strings are
independent and particles emitted in different strings are uncorrelated. This assumption works
well in $pp$, $pA$ and peripheral $AB$ collisions. Ho\-we\-ver, we shall see that for central $AB$
collisions the CERN data require the introduction of final state interaction of secondaries.
\par

Apart from this assumption the model has a sound theoretical basis and has great predictive
power. The latter is due to the fact that strings are universal. The momentum distribution
and fragmentation functions, allowing to compute the properties of each individual string, are
the same in all hadronic and nuclear collisions. Moreover, the momentum distribution functions
are determined from Regge intercepts. For instance, that of a valence quark in a pion is given by

\beq
\rho_{q_v}^{\pi}(x_{q_v}) \propto x_{q_v}^{-1/2} \ \left ( 1 - x_{q_v} \right )^{-1/2} \  .
\label{4ea}
 \eeq

\noindent The behaviour for $x \to 0$ is the well known $1/\sqrt{x}$ behaviour associated to an
ordinary Regge trajectory of intercept 1/2. Since $x_{q_v} + x_{\bar{q}_v} = 1$, the behaviour for
$x_{q_v} \to 1$ is the same as for $x_{\bar{q}_v} \to 0$, and therefore one has the symmetric
behaviour (\ref{4ea}). In this way the end point behaviour is under control. Of course, we can
multiply (\ref{4ea}) by a smooth function of $x$. However, in view of the singular behaviour at the
end points, this will have little influence on the numerical results. \par

For a valence quark in a proton, one has instead

\beq
\rho_{q_v}^p(x) \propto x^{-1/2} (1 - x)^{1.5} \ .
\label{3e}
\eeq

\noindent Here the behaviour at $x \to 1$ is obtained by ta\-king the diquark momentum fraction to
zero. This behaviour is controlled by an exotic $qq\bar{q}\bar{q}$ Regge trajectory of intercept
close to - 3/2. As a consequence of (\ref{3e}) the valence quark in a proton is slow in average
whereas the diquark is fast $(x_{q_v} + x_{qq} = 1)$. Its fragmentation into a lea\-ding baryon
accounts, in this way, for the leading particle effect (see Section 2.3). \par

In the first applications of DPM there was no attempt to compute the fragmentation functions.
Only their universality was used. For instance in studying $pA$ collisions, one can obtain the
fragmentation function from a fit to $pp$ data (or the $pA$ data for a given $A$). The model
then predicts both the $A$ dependence and the energy dependence. An important progress in the
model happened when Kaidalov showed \cite{3r} that the fragmentation functions can also be
determined to a  large extent from Regge arguments. In this way the model becomes very
predictive and the spectra of different hadrons can be computed basically without free
parameters - except absolute normalizations. \par

In order to illustrate this point, I show in Figs. 3 and 4 the leading proton and leading $\Lambda$
$x_F$ distribution in $pp$ collisions \cite{4r}. The shape of the leading proton spectrum is
practically fixed theoretically (here we show only the non-diffractive part but the diffractive one
can also be computed in a si\-mi\-lar way \cite{5r}). Moreover, the different shapes of the proton
and $\Lambda$ distributions is entirely due to the difference between $\alpha_R (0) = 1/2$, which
appears in the power of $1 - x_F$ for the lea\-ding proton, and $\alpha_{\phi} (0) = 0$, which
controls the $x_F \to 1$ behaviour of the $\Lambda$-distribution. Likewise, the sharp decrease of
$\bar{p}$ and $\bar{\Lambda}$ distributions, also shown in Figs. 3 and 4, is entirely controlled by
known Regge intercepts \cite{4r}, \cite{5r}. 

\begin{figure}
\vspace{7 truecm}
\caption{Feynman $x$ distribution of proton and antiproton in $pp$ collisions at 200
GeV/c compared with the data of ref. [30] at 175 GeV/c. Diffraction is not included in
the calculation.}
\end{figure} 

\begin{figure}
\vspace{7 truecm}
\caption{Feynman $x$ distribution of $\Lambda$ and $\bar{\Lambda}$ produced in $pp$
collisions at 200 GeV/c are compared to a compilation of data from ref. [5].} 
\end{figure}

\subsection{Nucleus-nucleus collisions} 
The generalization of the model to nucleus-nucleus collisions is rather straightforward. For
simplicity I consider here $AA$ collisions. In the approximation of only two strings per
nucleon-nucleon collision, the rapidity distribution of se\-con\-da\-ries is given by

\[ {dN^{AA} \over dy} (y) = \bar{n}_A \left [ N^{qq_{A_P}-q_{A_T}^v}(y) + N^{q^v_{A_P}-qq_{A_T}}(y) \right ] 
 \]
\beq
 + 2 (\bar{n} - \bar{n}_A) \ N^{q_s-\bar{q}_s}(y) \ \ \ .
\label{4e} \eeq

\noindent Here $N(y)$ are the rapidity distributions of the individual strings, $\bar{n}_A$ is
the average number of wounded nucleons of $A$ and $\bar{n}$ is the average number of nucleon-nucleon collisions.
Both $\bar{n}_A$ and $\bar{n}$ can be computed in the Glauber model. For instance for an average
collision (i.e. integrated over impact parameter), one has

\beq
\bar{n} = A^2 \sigma_{NN}/\sigma_{AB} \propto A^{4/3} \ .
\label{5e}
\eeq

\noindent Note that the total number of strings is $2\bar{n}$, i.e. two strings per inelastic
nucleon-nucleon collision. \par

The interpretation of (\ref{4e}) is obvious. With $\bar{n}_A$ struck nucleons, we have at our
disposal $\bar{n}_A$ diquarks of projectile and target ($qq_{A_P}$ and $qq_{A_T}$, respectively)
and as many valence quarks. This accounts for the first term of (\ref{4e}). The re\-mai\-ning
strings~: $2(\bar{n} - \bar{n}_A)$ have to be stretched by sea quarks and antiquarks, because
the a\-vai\-la\-ble valence constituents are all included in the first term~; this accounts for the
second term of (\ref{4e}). Of course we should combine the valence and sea constituents of the
projectile with those of the target in all possible ways. However, for linear quantities such
as multiplicities, each ordering gives practically the same result. \par

We can see from (\ref{4e}) and (\ref{5e}) that, if all strings would have the same plateau
height (i.e. the same value of $N(0)$), the plateau height in an average $AA$ collision would
increase like $A^{4/3}$. However, at present energies, the plateau height of the
$q_s$-$\bar{q}_s$, strings is smaller than that of $qq$-$q$ ones, and the first term of
(\ref{4e}) dominates. At higher energies the contribution of the sea strings becomes
increasingly important. Therefore, in order to make predictions for RHIC and LHC we have to
introduce the multistring configurations in each nucleon-nucleon collision. If their average
number is $2\bar{K}$ (this number can be computed in a generalized eikonal model~; one gets
$\bar{K} \simeq 2$ at $\sqrt{s} = 200$ and $\bar{K} \approx 3$ at $\sqrt{s} = 7$ TeV) the total
number of strings is $2\bar{K}\bar{n}$, and eq. (\ref{4e}) is changed into

\[
{dN^{AA} \over dy} (y) = \bar{n}_A \left [ N^{qq_{A_P}-q_{A_T}^v}(y) + N^{q_{A_P}^v-qq_{A_T}}(y)
\right .\]
\beq
\left . + (2 \bar{K} - 2) N^{q_s-\bar{q}_s} \right ] + (\bar{n} - \bar{n}_A) 2\bar{K}
N^{q_s-\bar{q}_s} \ . \label{6e} \eeq      

\noindent The predictions \cite{6r} for average and central ($b \approx 0$) $SS$ and $Pb$-$Pb$
collisions at RHIC and LHC energies are given in Table 1.  

\begin{table}
\caption{The number of charged secondaries per unit rapidity at $y^* = 0$ for average
($\rho_0^{aver}$) and central collision ($\rho_0^{central}$) at RHIC and LHC [6].} 
\begin{tabular}{cccc}
\hline
&$\sqrt{s}$ &$\rho_0^{aver.}$ &$\rho_0^{central}$ \\
\hline
SS &200 GeV &45 &166 \\
SS &7 TeV &134 &503 \\
Pb Pb &200 GeV &513 &2030 \\
Pb Pb &7 TeV &1890 &7900 \\
\hline
\end{tabular} 
\end{table}  

\subsection{Diquark breaking and the leading baryon} 

>From Eq. (\ref{3e}) it follows that the diquark is fast in average and carries a large longitudinal
fraction of the incoming proton. Its fragmentation will produce a leading baryon as illustrated in
Figs. 1 and 2. These configurations cor\-res\-pond to the fragmentation of the diquark as a whole
(the two valence quarks of the diquark find themselves into the same final state baryon), and will
be called in what follows diquark pre\-ser\-ving (DP) components. However, the diquark can also break
in the way illustrated in Figs. 5 and 6, which we shall call in what follows diquark brea\-king (DB)
components. It was stressed in \cite{7r} that these two components give contributions to the
central plateau height which have different e\-ner\-gy behaviour. Therefore when studying $pp$
collisions in a broad range of energies, one should keep track separately of these two components.
This is even more necessary in nuclear collisions since the two components have very different
$A$-dependence (the ratio of $DB$ over the $DP$ components increases with $A$). This is very
important since the $DB$ component produces baryons which are slower than those produced by the
$DP$ one. 
\begin{figure}
\vspace{6 truecm}
\caption{Same as Fig. 2 for the diquark brea\-king component.}
\end{figure}

\begin{figure}
\vspace{7 truecm}
\caption{Same as Fig. 5 but here the pro\-du\-ced baryon is formed by three sea quarks.}
\end{figure}

In order to simplify the presentation I discuss $pA$ collisions. Following \cite{8r}, I divide the
total nucleon-nucleon cross-section into a $DP$ piece and a $DB$ one : $\sigma_{in} = \sigma_{DP}
+ \sigma_{DB}$. I assume that once the diquark has been destroyed in a collision with one nucleon
of the nucleus it cannot be reconstructed in further collisions with other nucleons of the
nucleus. The $NA$ cross-section involving $n$ inelastic $NN$ collisions is then given by 

\[
\sigma_{DB,n}^{NA}(b) = {A \choose n} \sum_{i=1}^n {n \choose i} \sigma_{DB}^i \
\sigma_{DP}^{n-i} \ T_A^n(b)  \]
\beq
\times \left [ 1 - \sigma_{in}\ T_A(b) \right ]^{A-n}\ .  \label{7e}
\eeq

Here $T_A(b)$ is the standard nuclear profile function at impact parameter $b$, normalized to
unity. In Eq. (\ref{7e}) we have replaced the usual factor $\sigma_{in}^nT_A^n$ corresponding to
the cross-section for $n$ inelastic collisions by the product $\sum\limits_{i=1}^n {n \choose i}
\sigma_{DB}^i \sigma_{DP}^{n-1} T_A^n = (\sigma_{in}^n - \sigma_{DP}^n)T_A^n(b)$. Indeed only the
term $\sigma_{DP}^nT_A^n$ will contribute to the diquark preserving cross-section. Summing in $n$ we
have

\beq
\sigma_{DB}^{NA}(b) = \sum_{n=1}^A \sigma_{DB,n}^{NA}(b) = 1 - \left [ 1 -
\sigma_{DB}T_A(b) \right ]^A \ .\label{8e} \eeq

\begin{figure}
\vspace{7 truecm}
\caption{Rapidity distribution of the $p$-$\bar{p}$ diffe\-rence in peripheral and central
$SS$ collisions (solid curves) [8]. The dashed curves are obtained with $\sigma_{DB} = 0$.
The data are from [11]. The dotted curve is our prediction for the nucleon minus antinucleon
rapidity distribution in central $Pb$-$Pb$ collisions.} \end{figure}

\noindent Eq. (\ref{8e}) shows that the diquark preserving cross-section belongs to a class of
processes \cite{9r} which has only self-absorption (or self-shadowing). Obviously 

\beq
\sigma_{DP}^{NA}(b) = \sigma_{in}^{NA} (b) - \sigma_{DB}^{nA}(b) \ .
\label{9e}
\eeq

Since $\sigma_{DB} < \sigma_{in}$, it is clear from Eqs. (\ref{8e}) and (\ref{9e}) that
$\sigma_{DB}^{NA}$ increases with $A$ faster than $\sigma_{DP}^{NA}$. Actually, when $\sigma_{DB}$
is sufficiently small to ne\-glect in Eq. (\ref{8e}) second and higher powers of $\sigma_{DB}$,
$\sigma_{DB}^{NA}$ will increase linearly with $A$. This proves the result stated above that the
relative size of the $DB$ component increases with increasing $A$. The result can be easily
generalized to an $AB$ collisions. We have \cite{10r}

\beq
\sigma_{DB}^{AB}(b) = 1 - (1 - \sigma_{DB} T_{AB}(b))^{AB} 
\label{10e}
\eeq
\noindent where $T_{AB}({\bf b}) = \int d^2s T_A({\bf s}) T_B({\bf b} - {\bf s})$. For $\sigma_{DB}$
sufficiently small we get from (\ref{10e}), after integration in impact parameter, $\sigma_{DB}^{AB}
= AB \sigma_{DB}$. \par

In the numerical calculations, I take $\sigma_{DB}$ = 7 mb corresponding to a 20 $\%$ weight of the
$DB$ component. Using eq. (\ref{10e}) one finds that this weight has increased to 40 $\%$ for
central $SS$ collisions and is as large as 64 $\%$ in central $Pb$-$Pb$ collisions. \par

The results for peripheral and central $SS$ collisions at CERN energies are shown in Fig. 7 and
compared with the data from the experiment NA35 \cite{11r}. We also show the results obtained with
$\sigma_{DB} = 0$. We see that the dramatic dip present in the latter case for central $SS$
collision has been largely filled in by the contribution of the $DB$ component in agreement with
experiment. The prediction for a central $Pb$-$Pb$ collision is also shown. In this case the dip is
converted into a broad plateau - in agreement with pre\-li\-mi\-na\-ry data from the NA49
collaboration \cite{12r}.

\subsection{DPM and QGP signals}

In the following I will discuss a possible interpretation of two QGP signals (strangeness
enhancement and $J/\psi$ suppression) in the framework of DPM as well as the required
modifications of the model - such as string fusion and final state interaction of secondaries. I
will also discuss possible consequences of string fusion for cosmic ray physics. \par

\subsection{Strangeness enhancement}

How can one explain in DPM the enhancement of the ratios (\ref{3ea})~? Since the average number of
strings increases with $A$ (or with centrality), and since the extra strings involve sea quarks
and antiquarks at their ends, strangeness will be enhanced provided the fraction of $s$-quarks in
the sea is larger than the ratio of strange over non-strange particles at present energies.
However, this mechanism is numerically important only for kaons. $\Lambda/\bar{\Lambda}$
production in a $q_s$-$\bar{q}_s$ string is ne\-gli\-gea\-ble at present energies due to threshold
effects. A possible way to enhance $\Lambda/\bar{\Lambda}$ production is via string fusion. When
two strings overlap, both in transverse space and in rapidity, their momenta as well as their
quantum number can merge. If for instance two strings $u\bar{s}$ and $s\bar{d}$ fuse into a
$us$-$\bar{s}\bar{d}$ one, it will be easy to produce a $\Lambda/\bar{\Lambda}$ pair. In this way it
has been possible to explain $\bar{\Lambda}$ enhancement \cite{13r}. \par

While this is a very plausible mechanism, the evidence for string fusion is not yet compelling.
Indeed, one can assume that diquark-antidiquark pairs are present in the proton sea (with the
same relative amount required in the string breaking process in order to produce baryon pairs)
\cite{14r}. This mechanism is equivalent to string fusion in order to explain $\bar{\Lambda}$
enhancement but does not have some of the consequences of string fusion that will be discussed later.
\par

A limitation of the string fusion mechanism is that it gives the same enhancement of $\Lambda$
and $\bar{\Lambda}$ in absolute value. Experimentally the enhancement of $\Lambda$ is much
bigger. What can produce it~? First, the $DB$ component introduced in Section 2.3 is a source of
$\Lambda$-enhancement at mid-rapidities. Indeed, in the $DB$ diagram of Fig. 6 the
pro\-ba\-bi\-li\-ty to produce a $\Lambda$ is three times larger than in the $DP$ component of Fig. 2
since the strange quark can be any of the three sea quarks that form the baryon. For the $DB$
diagram of Fig. 5 this probability is two times larger than that of the $DP$ component. However,
the calculation shows that the obtained $\Lambda$-enhancement is substantially smaller than the
experimental one \cite{15r}. Another source of $\Lambda$ enhancement is therefore needed. Such a
source is provided by final state interaction of the produced secondaries. The most important
interactions turn out to be \cite{4r,15r,16r}

\beq
\pi + N \to K + \Lambda (K^* + \Lambda , \cdots )
\label{11e}
\eeq

\beq
\pi + \bar{N} \to K + \bar{\Lambda} (K^* + \bar{\Lambda} , \cdots ) \ . \label{12e}
\eeq
\noindent Since the strange particle yield is proportional to the product of densities of the
interacting particles, it is clear that $\Lambda$ enhancement will be more important than that of
$\bar{\Lambda}$ due to the larger density of nucleons as compared to antinucleons. Moreover, the
reactions

\beq
\pi + \Lambda \to K + N \quad , \quad \pi + \bar{\Lambda} \to K + \bar{N} \ , \label{13e}
\eeq 

\noindent which produce a decrease of $\Lambda$ and $\bar{\Lambda}$, are less important than
(\ref{11e})-(\ref{12e}) due to the smaller density of $\Lambda$ ($\bar{\Lambda}$) as compared to
nucleons (antinucleons). \par

Calculations along these lines have been performed in \cite{4r,15r,16r}. The results
for central $SS$ collisions are given in Table 2 and compared with experiment \cite{17r}. The
agreement is reasonably good. We see that final state interaction is needed in order to reach
agreement with experiment. Predictions for central $Pb$ $Pb$ collisions are given in Table 3. The
value with final state interaction is consistent with preliminary data from the NA49 collaboration
\cite{18r} (23 $\pm$ 7 at $y^* \sim$  0).

\begin{table}
\caption{$\Lambda$ rapidity distribution in central $SS$ collisions at 200 GeV/c per nucleon
computed with and without final state interaction (FSI) compared to data from the NA35
collaboration [17].} 
\begin{tabular}{cccc}
\hline
$y^*$ &No FSI &With FSI &NA35 \\
\hline
0 &0.78 &1.9 &2.2 $\pm$ 0.3 \\
0.5 &0.77 &1.8 &2.1 $\pm$ 0.3 \\
1 &0.75 &1.7 &2.1 $\pm$ 0.3 \\
1.5 &0.70 &1.6 &2.2 $\pm$ 0.3 \\
2 &0.62 &1.3 &1.4 $\pm$ 0.3 \\
\hline
\end{tabular}
\end{table}

\begin{table}
\caption{$\Lambda$ rapidity distribution in central $Pb$ $Pb$ collisions at 160 GeV/c per nucleon
computed with and without final state interaction (FSI).}
\begin{tabular}{ccc}
\hline
$y^*$ &No FSI &With FSI \\
\hline
0 &8.4 &23 $\div$ 31 \\
0.5 &8.1 &22 $\div$ 30 \\
1 &6.8 &20 $\div$ 25 \\
1.5 &5.0 &16 $\div$ 19 \\
2 &2.9 &8.8 $\div$ 9.1 \\
\hline
\end{tabular}
\end{table}

\subsection{J/$\bfpsi$ suppression}

We turn next to $J/\psi$ suppression \cite{19r}. For all $pA$ and $AB$ data \cite{20r} involving a
light projectile (up to $S$) there is an alternative explanation to Debye color screening
proposed in \cite{19r}. This explanation is based on the following me\-cha\-nism \cite{21r}. In the
present experimental conditions the $J/\psi$ is produced outside (behind) the nucleus. Ho\-we\-ver,
the $c\bar{c}$ pair is produced inside the nucleus and can interact with nucleons in its path through
the nucleus. This interaction can modify its wave function in such a way that it has a vanishing
projection into the $J/\psi$ bound state. A single absorptive cross-section, $\sigma_{abs}
\approx$ 6.2 mb, allows to describe all data except the $Pb$ $Pb$ ones. This is illustrated
in Fig. 8. The data \cite{22r} for $Pb$ $Pb$, also displayed in this figure, show clearly that
nuclear absorption alone does not work. A\-no\-ther ``conventional'' mechanism is the final state
interaction of the $J/\psi$ bound state with other secondaries produced at comparable velocity
(co-movers) producing open charm~: \par

\begin{figure}
\vspace{8 truecm}
\caption{The ratio $B_{\mu \mu} \sigma (J/\psi )/\sigma (DY)$ versus the interaction
length $L$ for $pp$, $pA$, $SU$, and $Pb$ $Pb$ interactions. Data are from refs. [20,22]. The
straight line is obtained with nuclear absorption alone. The points labelled theory are obtained
[25] with nuclear absorption plus final state interaction of the $J/\psi$ bound state.}
\end{figure} 

\beq
\pi + J/\psi \to D + \bar{D} + \cdots 
\label{14e}
\eeq  

\noindent This  idea was introduced in the literature a long time ago
\cite{23r}. However, recent theoretical calculations have shown that the $\pi + J/\psi$ and $N +
J/\psi$ cross-sections increase very slowly from threshold (in contrast to $\psi ' + \pi$ which
has a very rapid increase) \cite{24r}. Therefore in the case of the $J/\psi$ the co-mover
cross-section is expected to be very small (not larger than 0.5 mb). Because of that the idea of
$J/\psi$ suppression via interaction with co-movers has been progressively abandoned. \par

In a recent paper \cite{25r} we have shown that introducing the final state interaction (\ref{14e})
with a cross-section of 0.4 mb, one can explain the recent $Pb$ $Pb$ data. The results are shown in
Fig. 8. A similar result has been obtained independently in \cite{26r}. \par

One should note, however, that when in\-tro\-du\-cing the interaction with co-movers one has to
reduce the value of the cross-section for nuclear absorption. In this way the $J/\psi$ yield in $pA$
collisions does not fall with increasing $A$ as steeply as the data seem to indicate (the
``effective'' absorptive cross-section resulting from the combined effect of nuclear absorption and
in\-te\-rac\-tion with co-movers is 5 mb while experimental data require 6.2 $\pm$ 0.7 mb). Moreover
the $J/\psi$ yield in $SU$ falls somewhat faster than indicated by experiment. At present, however,
the data are not precise enough to rule out such an explanation and further study is needed to
distinguish it from a QGP scenario \cite{27r} and/or other collective effects such as string
percolation \cite{28r}. \par

There are, of course, many other interesting results obtained at CERN which I have not mentioned.
In particular Bose-Einstein correlation measurements (interferometry) indicate that the size of
the interaction region is larger than the geo\-me\-tri\-cal size of the colliding system. Again, this
requires the existence of final state interaction of secondaries. 

\subsection{Consequence of string fusion for cosmic ray physics}

As discussed in section 2.5, there are hints in the data in favor of string fusion, although
there is no compelling evidence yet. Nevertheless, it is interesting to examine its possible
consequences in cosmic ray physics. In a recent paper the Santiago group \cite{29r} discusses two
such consequences~: the rise of the average shower depth of maximum for cosmic rays in the energy
range $10^{16}$ to $10^{19}$ eV and the possibility that string fusion acts as a hadronic
accelerator. \par

When the energy increases, string fusion becomes more important and produces a decrease of the
average multiplicity. Taking a fixed chemical composition (i.e. independent of energy), consisting of
90~$\%$ of iron and 10~$\%$ of protons, one gets an average multiplicity, which, in the range
$10^{17}$ to $10^{19}$~eV, is identical to the one obtained without string fusion and a chemical
composition of primaries changing with energy (from 75~$\%$ Fe and 25~$\%$ p at $10^{16}$ eV to
50~$\%$ Fe and 50~$\%$ p at $10^{19}$ eV). This is shown in Fig. 9. 

\begin{figure}
\vspace{7 truecm}
\caption{Total average multiplicity versus primary energy for a fixed composition
assuming string fusion (solid line) and with no string fusion and a composition that changes with
energy (dashed line). See text for details.} \end{figure}

Another interesting consequence of string fusion is that it produces secondaries outside the
kinematical range of a nucleon-nucleon in\-te\-rac\-tion. The authors can explain in this way the
so-called cumulative effect. Moreover, they claim that these two effects
can change the profile of extensive air showers and in this way alter
conclusions on the primary composition and energy. For example, provided
their composition could be determined by an independent method,
cosmic ray events with energies around
3.10$^{20}$ eV would contain, in the presence of string fusion, particles with $|x_F| > 2$ or 3 (or
more in the pre\-sen\-ce of string percolation \cite{28r}), and would thus cor\-res\-pond to an
energy for the primary upto $2\div 4$ times smaller than the one without string fusion. Such a
reduction in e\-ner\-gy could make these events compatible with the cut-off due to the scattering of
cosmic rays with the microwave background.

\section{CONCLUSIONS}

The QGP search has gained full credibility with the lattice QCD results. Experimental data are
of very high quality but give only some hints of QGP formation. Future data and especially
future machines will be decisive in this search. \par

In the mean time, we have gained new insight in the dynamics of multiparticle production
allowing to improve current models (strong baryon stopping hinting at diquark breaking, string
fusion with interesting consequences for cosmic ray physics and final state interaction of
secondaries). \par

Finally, I want to stress once more that the CERN heavy ion program has produced very many
interesting results not mentioned here.


\begin{thebibliography}{11}  
\bibitem{1r} A. Capella, U. Sukhatme, C.-I. Tan and J. Tran Thanh Van, Phys. Rep. 236 (1994)
225.  
\bibitem{2r} A. Capella, Proceedings 22nd International Cosmic Ray Conference, Dublin, Ireland,
1991.   \bibitem{3r} A. B. Kaidalov, Yad. Fiz. 45 (1987) 1452.     
\bibitem{4r} A. Capella, A. Kaidalov, A. Kouider Akil, C. Merino and J. Tran Thanh Van, Z.
Phys. C70 (1996) 507.     
\bibitem{5r} A. B. Kaidalov and O. I. Piskunova, Sov. J. Nucl. Phys. 41 (1985) 816.  
\bibitem{6r} A. Capella, C. Merino and J. Tran Thanh Van, Phys. Lett. B265 (1991) 415. 
\bibitem{7r} B. Z. Kopeliovich and B. G. Zakharov, Z. Phys. C43 (1989) 241.
\bibitem{8r} A. Capella and B. Z. Kopeliovich, Phys. Lett. B381 (1996) 325.
\bibitem{9r} A. Blankenbecler, A. Capella, C. Pajares, A. V. Ramallo and J. Tran Thanh Van, Phys.
Lett. B107 (1981) 106. 
\bibitem{10r} C. Pajares and A. V. Ramallo, Phys. Rev. D31 (1985) 2800. 
\bibitem{11r} NA35 collaboration, H. Str\"obel et al., Nucl. Phys. A525 (1991) 59c.
\bibitem{12r} NA49 collaboration in Proceedings XXXI Rencontres de Moriond, Les Arcs (France), 1996
ed. J. Tran Thanh Van~; see also NA44 collaboration, ibid. 
\bibitem{13r} N. Armesto, M.A. Braun, E. G. Ferreiro and C. Pajares,
Phys. Lett. B344 (1995) 301. References to earlier papers on string fusion can be found in G.
Gustafson, Nucl. Phys. A566 (1994) 233c.  RQMD~: H. Sorge, R. Matiello, A. von Keetz, H.
St\"ocker, W. Greiner, Z. Phys. C47 (1990) 629~; H. Sorge, M. Berenguer, H. St\"ocker, W.
Greiner, Phys. Lett. B289 (1992) 6~; Th. Sch\"onfeld et al., Nucl. Phys. A544 (1992) 439c.
\bibitem{14r} J. Ranft, A. Capella, J. Tran Thanh Van, Phys. Lett. B320 (1994) 346~; H. J.
M\"ohring, J. Ranft, A. Capella, J. Tran Thanh Van, Phys. Rev. D47 (1993) 4146 (the calculations in
these papers are based on the DPMJET and DTNUC codes).
\bibitem{15r} A. Capella, Phys. Lett. B364 (1995) 175. 
\bibitem{16r} A. Capella, preprint LPTHE Orsay 96-30, hep-ph 96-05216, Phys. Lett. B in press. 
\bibitem{17r} NA35 collaboration, T. Alber et al, Z. Phys. C64 (1994) 195.
\bibitem{18r} NA49 collaboration, Proceedings QM 96, Heidelberg. 
\bibitem{19r} T. Matsui and H. Satz, Phys. Lett. B178 (1986) 416. 
\bibitem{20r} NA38 collaobration, C. Baglin et al., Phys. Lett. B201 (1989) 471~; Phys. Lett. B255
(1991) 459. 
\bibitem{21r} R. Salmeron, Nucl. Phys. A566 (1994) 199c. A.
Capella, J. A. Casado, C. Pajares, A.V. Ramallo and J. Tran Thanh Van, Phys. Lett. 
B206 (1988) 354. A. Capella, C. Merino, C. Pajares, A. V. Ramallo and J. Tran
Thanh Van, Phys. Lett. B230 (1989) 149.
C. Gerschel and J. H\"ufner, Phys. Lett. B207, 253 (1988)~; Z. Phys.
C56 (1992) 71. 
\bibitem{22r} NA50 collaboration, P. Bordal\'o et al., Proceedings XXXI Rencontres de Moriond 1996
ibid~. M. Gonin et al., Proceedings Quark Matter 96, ibid. 
\bibitem{23r} J.P. Blaizot and J.Y. Ollitrault, Phys. Rev. D39 (1989) 232.
S. Gavin, M. Gyulassy and A. Jackson, Phys. Lett. 207 (1988) 194.
J. Ftacnik, P. Lichard and J. Pitsut, Phys. Lett. B207 (1988) 194.
R. Vogt, M. Parakash, P. Koch and T. H. Hansson, Phys. Lett. B207
(1988) 263. S. Gavin and R. Vogt, Nucl. Phys. B345 (1990) 1104.
S. Gavin, H. Satz, R. Thews and R. Vogt, Z. Phys. C61 (1994) 351.
\bibitem{24r} G. Bhanot and M. E. Peskin, Nucl. Phys. B156 (1979) 365.
A. Kaidalov and P. Volkovitsky, Phys. Rev. Lett. 69 (1992) 3155.
A. Kaidalov, Proceedings XXVIII Rencontres de Moriond (1993), ed. J. Tran
Thanh Van. M. Luke et al, Phys. Lett. B288 (1992) 355.
D. Kharzeev and H. Satz, Phys. Lett. B306 (1994) 155.
\bibitem{25r} A. Capella, A. Kaidalov, A. Kouider Akil and C. Gerschel, LPTHE 96-55, hep-ph/9607265.
 \bibitem{26r} S. Gavin et R. Vogt, Proceedings QM 96, ibid and preprint LBL-37980 (1996).
\bibitem{27r} J. P. Blaizot and J. Y. Ollitrault, Pro\-cee\-dings Quark Matter 96, ibid; Phys. Rev. Lett. 77 (1996) 1703.
C. Y. Wong, Proceedings Quark Matter 96, ibid. D. Kharzeev and H. Satz,
Pro\-cee\-dings QM96, ibid.  \bibitem{28r} N. Armesto, M. A. Braun,
E. G. Ferreiro and C. Pajares, Phys. Rev. Lett. 77 (1996) 3736.
\bibitem{29r} N. Armesto, M. A. Braun, E. G. Ferreiro, C. Pajares and Yu M.
Shabelski, preprints US-FT/6-96 (Phys. Lett. B to appear) and US-FT/11-96
(Astropart. Phys. to appear). \bibitem{30r} A. E. Brenner et al., Phys. Rev. D26 (1982) 1497.
\end{thebibliography}
\end{document}